\documentclass[a4paper]{jpconf}
\usepackage{graphicx}
\usepackage{cite}
\begin{document}
\title{Two-dimensional macroscopic quantum tunneling in multi-gap superconductor Josephson junctions}

\author{Hidehiro Asai$^1$, Shiro Kawabata$^1$, Yukihiro Ota$^2$, Masahiko Machida$^2$}

\address{$^1$Electronics and Photonics Research Institute (ESPRIT), National Institute of Advanced Industrial Science and Technology(AIST), Tsukuba, Ibaraki 305-8568, Japan

$^2$CCSE, Japan Atomic Energy Agency, Kashiwa, Chiba 277-8587, Japan
}

\ead{hd-asai@aist.go.jp}

\begin{abstract}
Low-temperature characters of superconducting devices yield definite
 probes for different superconducting phenomena.  
We study the macroscopic quantum tunneling (MQT) in a Josephson
 junction, composed of a single-gap superconductor and a two-gap
 superconductor. 
Since this junction has two kinds to the superconducting phase
 differences, calculating the MQT escape rate requires the analysis of quantum
 tunneling in a multi-dimensional configuration space. 
Our approach is the semi-classical approximation along a 1D curve in a 2D
 potential-energy landscape, connecting two adjacent potential (local)
 minimums through a saddle point. 
We find that this system has two plausible tunneling paths; an in-phase
 path and an out-of-phase path. 
The former is characterized by the Josephson-plasma frequency, whereas
 the latter is by the frequency of the characteristic collective mode in
 a two-band superconductor, Josephson-Leggett mode. 
Depending on external bias current and inter-band Josephson-coupling
 energy, one of them mainly contributes to the MQT. 
Our numerical calculations show that the difference between the in-phase
 path and the out-of-phase path is manifest, with respect to the
 bias-current-dependence of the MQT escape rate. 
This result suggests that our MQT setting be an indicator of the
 Josephson-Leggett mode. 
%
%
%
\end{abstract}

\section{Introduction}
Macroscopic quantum tunneling (MQT) is one of the interesting phenomena of
Josephson junctions (JJs)~\cite{CLMQT,Simanek}. 
A superconducting-to-resistive switching event in a current-biased JJ is
related to the MQT at low temperatures (i.e., negligible
contributions from thermal excitations). 
Different JJs show the tunneling
phenomena~\cite{1stMQT,EMMQT,Inomata;Kawabata:2005,Jin;Meuller:2006,Bauch;Lombardi:2006,Kubo;Takano:2012}. 
The theoretical aspects are studied well, depending on types of
the JJs~\cite{Kato;Imada:1996,Kawabata;Tanaka:2005,Machida;Koyama:2007,Savelev;Nori:2007,Sbochakov;Nori:2007,KawabataMQT,2gapMQTOta,HekkingMQT;2012,2gapMQTAsai}.


Subsequent to the discovery of magnesium diboride~\cite{1stMgb2} and
 iron-based superconductors~\cite{1stIron}, multi-band superconductivity
 has been extensively
 studied~\cite{IronReview,IronReview2,Mgb2Cp,Mgb2Review,Kito:2008}.  
A lot of notable effects have been predicted in the JJs with multi-band superconductors~\cite{2gapAgter,Onari;Tanaka:2009,Linder;Sperstad;Sudbo:2009,Inotani;Ohashi:2009,2gapOtaPRL,Golubov;Dolgov:2009}. 
In such JJs, the switching process out of the zero-voltage state is
 described by the dynamics of multiple gauge-invariant phase differences.
Hence, the MQT may occur in a multi-dimensional configuration space,
 defined by the phase differences. 
%
 
In this paper, we study the MQT in a hetero JJ, formed by a conventional
 single-gap superconductor and a two-gap superconductor. 
The switching events in such JJs take place in 2D configuration
 spaces, including the two superconducting phase differences. 
We calculate the MQT escape rate, taking the quantum
 phase dynamics in the 2D potential. 
This potential-energy landscape leads to the
 presence of two plausible tunneling paths; an in-phase path and an
 out-of phase path. 
The former is characterized by the Josephson-plasma frequency, whereas
the latter is by the frequency of the characteristic collective mode in
a two-band superconductor, Josephson-Leggett mode. 
In contrast to the in-phase path (see, e.g.,
 \cite{2gapMQTOta,2gapMQTAsai}), the contribution from the out-of-phase
 path to the MQT is rarely studied. 
Depending on external bias current and inter-band Josephson-coupling
 energy, one of them mainly contributes to the MQT. 
The in-phase path is primary for a strong inter-band coupling, while the
 out-of-phase path mainly contributes to the tunneling for a weak
 inter-band coupling. 
Our numerical calculations show that the difference between the in-phase
 path and the out-of-phase path is manifest, with respect to the
 bias-current-dependence of the MQT escape rate. 
In comparison with the MQT rate for the in-phase path, 
the MQT rate for the out-of-phase path drastically decreases as the bias current decreases.
Hence, by systematically measuring the MQT rate,
we can detect the tunneling process corresponding to the out-of-phase path, 
 which is related to the Josephson-Leggett mode. 
Thus, our MQT setting would be an indicator of the
 Josephson-Leggett mode. 
\section{Model and formulation}

\begin{figure}[h]
\includegraphics[width=18pc]{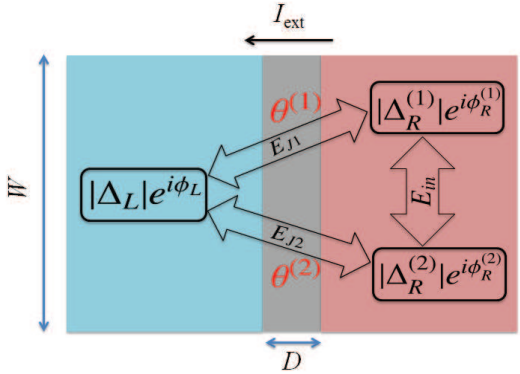}\hspace{2pc}%
\begin{minipage}[b]{18pc}\caption{\label{label}A schematic figure of the Josephson junction
 composed of a single-gap and a two-gap superconductor.}
\end{minipage}
\end{figure}

We consider a hetero Josephson junction composed of a single-gap superconductor and a two-gap superconductor, as shown in figure 1.
%
The superconducting electrodes are divided by an insulating layer with
thickness $D$ and area $W$.
The DC bias current $I_{\scriptsize \textrm{ext}}$ is applied to the junction.
The right electrode has two gaps, 
$|\Delta^{(1)}_{\rm R}| e^{i \phi^{(1)}_{\rm R}}$ 
and $|\Delta^{(2)}_{\rm R}| e^{i \phi^{(2)}_{\rm R}}$, 
whereas the left electrode has a single gap 
$|\Delta_{\rm L}| e^{i \phi_{\rm L}}$. 
These superconducting phases are coupled to each other via the
Josephson couplings,
and the two gauge-invariant phase differences, $\theta^{(1)}$ and
$\theta^{(2)}$ can be defined between the electrodes. 
%
%
Using these phase differences, the two important variables for describing MQT, i.~e.~, 
 the center-of-mass phase $\theta$ and relative phase $\psi$
are described as,
\begin{eqnarray}
\theta 
=
\frac{\alpha_2}{\alpha_1 + \alpha_2} \theta^{(1)} 
+ \frac{\alpha_1}{\alpha_1 + \alpha_2} \theta^{(2)},
\ \ \ \psi = \theta^{(1)} - \theta^{(2)}.
\end{eqnarray}
A dimensionless constant $\alpha_{i}$ is related to the density of
states of the $i$th-band electron near the interface between the right
electrode and the insulator.
 The Josephson relation in this junction is given by 
$
\frac{\partial \theta}{\partial t}
=
\frac{2e}{\hbar} V
$
\cite{2gapOtaPRL}
with the electric charge $e$, the Planck constant $\hbar$. 
Hence, $\theta$ is directly coupled with the electric field.

The real-time effective Lagrangian for the JJ is~\cite{2gapOtaPRL,2gapMQTOta,2gapMQTAsai}
\begin{eqnarray}
\mathcal{L} (\theta, \Psi)
&=& \frac{M}{2} \dot{\theta}^2 + \frac{M}{2} \dot{\Psi}^2 - V_{{\scriptsize \textrm{2D}}}(\theta, \psi), 
 \ \ \  M = \frac{\hbar^2}{2 E_{{\scriptsize \textrm{c}}}} \label{2DLagrangean}  \\
V_{{\scriptsize \textrm{2D}}}(\theta, \Psi) 
&=& -E_{{\scriptsize \textrm{J,1}}} \cos{\Bigl(\theta + \frac{\alpha_1}{r} \Psi \Bigr)} 
- E_{{\scriptsize \textrm{J,2}}} \cos{\Bigl(\theta - \frac{\alpha_2}{r} \Psi \Bigr)} 
- E_{{\scriptsize \textrm{in}}} \cos{\bigl(r \Psi \bigr)}
- E_{{\scriptsize \textrm{J}}} \gamma \theta, \label{2Dpotential}
\end{eqnarray} 
with the charging energy $E_{\textrm{c}}$, the Josephson energy in the $i$th
tunneling channel $E_{{\scriptsize \textrm{J},i}}$, the total Josephson energy
$E_{{\scriptsize \textrm{J}}}=E_{{\scriptsize \textrm{J},1}}+E_{{\scriptsize \textrm{J},2}}$, the inter-band Josephson energy 
\(
E_{{\scriptsize \textrm{in}}} = \kappa (\hbar W |J_{{\scriptsize \textrm{in}}}| /2e)
\), 
the dimensionless bias current $\gamma = I_{\scriptsize \textrm{ext}}/I_{\scriptsize \textrm{c}}$. 
Here, we use the scaled relative phase $\Psi = \psi/r, \ \ r = \sqrt{\alpha_1 + \alpha_2}$ in order that the mass of the relative phase equals to that of center of mass phase, $M$.
The critical current $I_{\scriptsize \textrm{c}}$ is related to $E_{\scriptsize \textrm{J}}$. 
The factor $\kappa$ in $E_{\scriptsize \textrm{in}}$ is the sign of the inter-band Josephson
coupling constant $J_{\scriptsize \textrm{in}}$, which is related to the gap symmetry. 
We take $\kappa=+1$ in this paper. 
 
As seen in (\ref{2Dpotential}), the potential for a fictitious phase particle is described in 2D ($\theta, \Psi$) space.
In this study, we calculate the MQT rate in the 2D space in the following manner.
First, we find a local minimum of the potential and the saddle point which is close to the local minimum.  
Next, by assuming that the fictitious particle follows the path of minimal curvature \cite{HekkingMQT;2012},
 we numerically calculate the tunneling path passing through the saddle point.
Finally, we calculate the MQT rate for tunneling barrier $V_{{\scriptsize \textrm{1D}}} (z)$ along the tunneling path
using Wentzel-Kramers-Brillouin (WKB) formula, where $z$ is the curvilinear abscissa along the path.
In the WKB approximation, the MQT rate $\Gamma$ is given by,~\cite{Simanek}
\begin{eqnarray}
\Gamma
= \omega_0 \bigl( \frac{30 S_{{\scriptsize \textrm{b}}}}{\pi \hbar} \bigr)^{\frac{1}{2}} \exp{\Bigl(- \frac{S_{{\scriptsize \textrm{b}}}}{\hbar}\Bigr)}, 
\ \ \ 
S_{{\scriptsize \textrm{b}}} = 2 \int^{\zeta_1}_{\zeta_0} d\zeta [2 M V_{{\scriptsize \textrm{1D}}}(\zeta)]^{\frac{1}{2}},
\end{eqnarray} 
where $\zeta_0$ and $\zeta_1$ are the coordinates for the local minimum and the end point of the tunneling path,
and $\omega_0$ is the characteristic frequency along the tunneling path.
 
\section{Numerical results}

\begin{figure*}[h]
\begin{center}
\includegraphics[width=15cm]{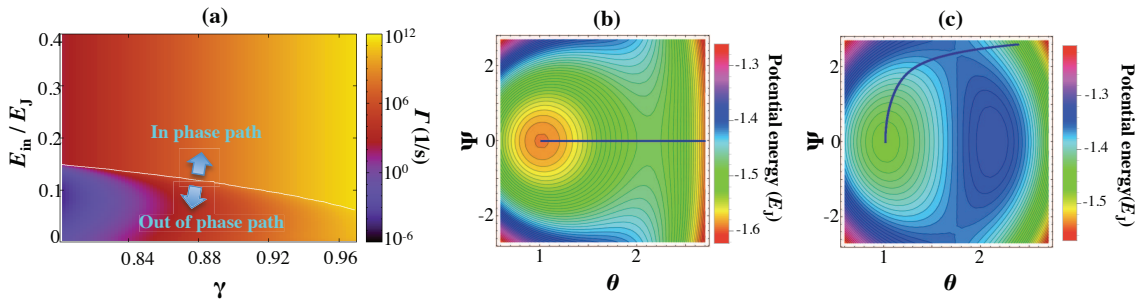}
\caption{\label{label} (a) MQT escape rate $\Gamma$ as functions of 
external current $\gamma$ and inter-band Josephson energy $E_{{\scriptsize \textrm{in}}}$. 
(b) and (c) are potential landscapes $V_{{\scriptsize \textrm{2D}}}(\theta, \Psi)$ 
 and tunneling paths in the landscapes
 for $E_{{\scriptsize \textrm{in}}}/E_{{\scriptsize \textrm{J}}}= 0.05, \gamma=0.85$ 
 and $E_{{\scriptsize \textrm{in}}} /E_{{\scriptsize \textrm{J}}}= 0.2, \gamma = 0.85$ respectively. 
Black solid lines indicates the tunneling paths. }
\end{center}
\end{figure*}

In order to investigate MQT for various tunneling path, 
we calculate the MQT rate $\Gamma$, 
varying the parameter sets $(E_{{\scriptsize \textrm{in}}}, \gamma$). 
In this study, we take 
$E_{{\scriptsize \textrm{J}}} = 0.01 \ \textrm{eV}, 
E_{{\scriptsize \textrm{c}}}/E_{{\scriptsize \textrm{J}}}= 0.002, 
\ E_{{\scriptsize \textrm{J,1}}} = E_{{\scriptsize \textrm{J,2}}}, 
\ \alpha_1 = \alpha_2=0.1$.
 Figure 2 (a) shows $\Gamma$ as functions of $E_{{\scriptsize \textrm{in}}}$ and $\gamma$. 
In this calculation, we find that two types of tunneling paths appear in our junction: 
in-phase and out-of-phase path.
The white solid line in figure 2(a) indicates the boundary between the parameter region corresponding to each tunneling path.
In figures 2 (b) and (c),
 we show the potential landscapes $V_{{\scriptsize \textrm{2D}}}(\theta, \Psi)$ 
 and the tunneling paths in the landscapes 
 for ($E_{{\scriptsize \textrm{in}}}= 0.05, \gamma=0.85$) 
 and ($E_{{\scriptsize \textrm{in}}} = 0.2, \gamma = 0.85$) respectively.
The in-phase path is the straight-line path along the $\theta$ axis with $\Psi=0$ as shown in figure 2 (b).      
Thus, $\omega_0$  corresponds to the Josephson plasma frequency,
 $\omega_0 = \sqrt{2 E_{{\scriptsize \textrm{J}}} E_{{\scriptsize \textrm{c}}} } (1-\gamma^2)^{\frac{1}{4}} /\hbar$. 
On the other hand, the out-of-phase path is the curved-line path with $\Psi \neq 0$. 
Moreover, in the vicinity of the local minimum, the path is almost parallel to the $\Psi$ axis.
Thus, $\omega_0$ corresponds to the Josephson Leggett frequency,
 $\omega_0 = \sqrt{r^2 E_{{\scriptsize \textrm{in}}} E_{{\scriptsize \textrm{J}}}}/ \hbar$.
 As shown in figure 2 (a),  out-of-phase path only appears in weak inter-band coupling condition
  $E_{{\scriptsize \textrm{in}}}/E_{{\scriptsize \textrm{J}}} < 0.15$.
Moreover,  
the MQT rate corresponding to the out-of-phase path drastically decreases as $\gamma$ decreases
compared to the in-phase path.
Figure 3 (a) and (b) show the escape rate $\Gamma$, respectively,  
for $E_{{\scriptsize \textrm{in}}}/E_{{\scriptsize \textrm{J}}} =0.08$ and $E_{{\scriptsize \textrm{in}}}/E_{{\scriptsize \textrm{J}}} =0.12$,
varying $\gamma$. 
The peculiar kink structures in the both figures are ascribed to switching from the out-of-phase to the in-phase tunneling paths, 
that is, the bias-current dependence of $\Gamma$ of the both paths are largely different.
%
We check that the kink position in figure 3(a) ($\gamma = 0.947$) is consistent with the data in figure 2(a);
the white solid line in figure 2(a) intersects with  $E_{{\scriptsize \textrm{in}}}/E_{{\scriptsize \textrm{J}}} =0.08$ at this current values.
A similar analysis is done for $E_{{\scriptsize \textrm{in}}}/E_{{\scriptsize \textrm{J}}} =0.12$, indicating that the kink appears at $\gamma = 0.877$.
%
%
Figure 4 shows the remarkable difference between the out-of-phase path (figure 4(a)) and the in-phase path (figure 4(b)),
on the potential landscape for $E_{{\scriptsize \textrm{in}}}/E_{{\scriptsize \textrm{J}}} =0.08$.
 %
%
 %
%
%
The MQT along the out-of-phase path reflects the nature of the inter-band fluctuation, namely, the Josephson Leggett mode.
Hence, we can detect the Josephson Leggett mode by systematically measuring the MQT rate for  multi-band JJs with weak inter-band coupling.
   
\begin{figure*}[h]
\begin{center}
\includegraphics[width=12cm]{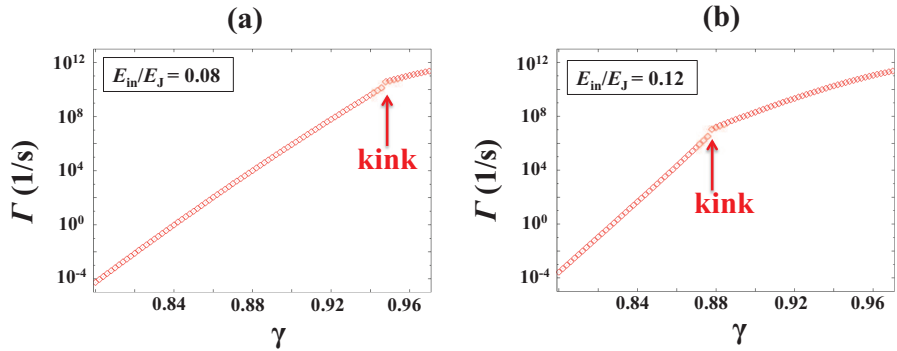}%
\caption{\label{label} 
MQT escape rate $\Gamma$ as a function external current $\gamma$ 
for (a) $E_{{\scriptsize \textrm{in}}}/E_{{\scriptsize \textrm{J}}}= 0.08$,
and (b) $E_{{\scriptsize \textrm{in}}}/E_{{\scriptsize \textrm{J}}}= 0.12$.
}
\vspace{-1cm}
\end{center}
\end{figure*}
\begin{figure*}[h]
\begin{center}
\includegraphics[width=12cm]{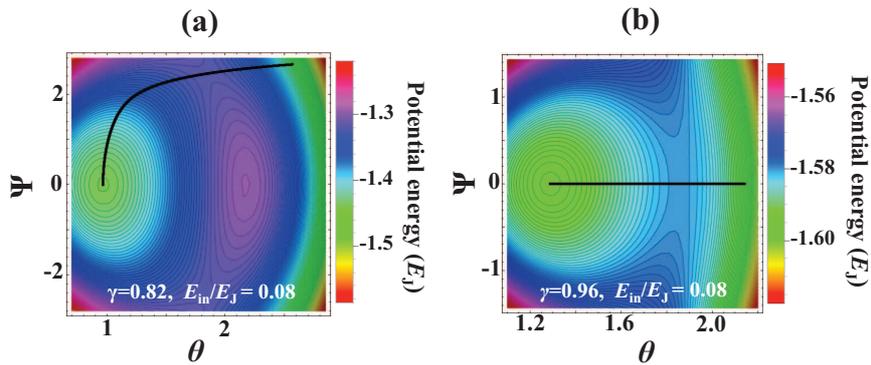}%
\caption{\label{label} 
Potential landscapes and tunneling paths in the landscapes for
(a) ($E_{{\scriptsize \textrm{in}}}/E_{{\scriptsize \textrm{J}}}= 0.08, \gamma = 0.82$),
and (b) ($E_{{\scriptsize \textrm{in}}}/E_{{\scriptsize \textrm{J}}}= 0.08, \gamma = 0.96$).
Black solid lines indicates the tunneling paths.  
}
\end{center}
\end{figure*}
%

\section{Conclusion}
We investigated the MQT in a hetero JJ 
consisting of a single-gap and a two-gap superconductor.
In this JJ, the  MQT can be described by quantum dynamics of two phase variables: center-of-mass phase $\theta$ and relative-phase $\psi$.
We analyzed the tunneling path in 2D potential-energy landscape for the phase dynamics,
 and calculated MQT rate along the path.
We found two possible tunneling paths in 2D potential-energy landscape, i.e., in-phase and out-of phase path. 
The in-phase path is the straight-line path along the $\theta$ axis with $\psi=0$, 
 and related to the Josephson plasma mode. 
On the other hand, the out-of-phase path is the curved-line path with $\psi \neq 0$, 
and related to the inter-band fluctuation (i.e. Josephson Leggett mode). 
The relevant tunneling path changes with the external bias current and the inter-band Josephson-coupling energy. 
We found that the in-phase path appears when the inter-band coupling energy is strong,
whereas, the out-of phase path appears when the inter-band coupling energy is weak.
 Moreover, in comparison with the in-phase path,
the MQT rate corresponding to the out-of-phase path drastically decreases as bias current  decreases.
 %
%
Our results indicate that we can detect the Josephson Leggett mode
by systematically measuring the MQT rates for various inter-band coupling conditions.



\end{document}